# Exceedance Probabilities for Large Earthquakes
## From DIY Local Earthquake Ensemble Nowcasting and Forecasting: Magnitude, Natural Time, and Calendar Time

John B Rundle[1,2,3], Ian Baughman[1], Andrea Donnellan[4,3], Lisa Grant Ludwig[5], Geoffrey Fox[6], Kazuyoshi Nanjo[7]

[1] University of California, Davis, CA
[2] Santa Fe Institute, Santa Fe, NM
[3] Jet Propulsion Laboratory, Pasadena, CA
[4] Purdue University, West Lafayette, IN
[5] University of California, Irvine, CA
[6] University of Virginia, Charlottesville, VA
[7]Shizuoka University, Japan

## Abstract

This paper focuses on the problem of anticipating the local occurrence of future large earthquakes. "Local" is defined as the probability of a large earthquake occurring with a defined circle of arbitrary radius surrounding a point of interest. The main (and for that matter, the only) assumption for all these works is that the Gutenberg-Richter (GR) magnitude-frequency relation holds. Here we describe a method for computing calendar time forecasts in a local area for large earthquakes of a target magnitude $M_T$ using a count small earthquakes $M_S < M_T$ in the area. Using the idea that the GR relation is valid throughout the surrounding region, we define an ensemble of earthquakes in larger surrounding regions to be used in computing the forecast. What follows is simple data mining. The method has significant skill, as defined by the Receiver Operating Characteristic (ROC) test, which improves as time since the last major earthquake increases. The probability is conditioned on the number of small earthquakes $n(t)$ that have occurred since the last large earthquake. The probability is computed directly as the Positive Predictive Value (PPV) associated with the ROC curve. The method is validated by comparison to the UCERF3 forecasts for the UCERF3-defined geographic boxes centered on Los Angeles and San Francisco. The method is then applied to a 125-KM radius circular area around Los Angeles, California, following the January 17, 1994 magnitude $M6.7$ Northridge earthquake, and short term forecasts (1 year and 5 year ) are computed.

# 1 Introduction

## 1.1 Earthquake Nowcasting

Previous papers have developed several techniques for earthquake nowcasting. The nowcasting approach is based on "natural time", which is defined as counts of small earthquakes [1-4]. Natural time is the time scale that is relevant to the system dynamics, so earthquake nowcasting uses natural time to track the progression of the fault system through its cycle of large earthquake activity. The term "nowcasting "is used in the same sense as for weather and economic nowcasting [5,6], tracking the current state of a system in the recent past, current time, and the near future. Methods to produce earthquake nowcasts are the subject of many previous papers [5-39].

The nowcasting method is applied to a small circle of arbitrary radius surrounding a point of interest, typically a city. Computing the probability then involves a data mining phase using an expanding series or ensemble of larger rectangular regions surrounding the circle. These regions then comprise the ensemble for the method. Each member of the ensemble contains a number of earthquake "cycles" which begin and end on earthquakes larger than a target magnitude $M_T$. Each cycle contains a number of "small" earthquakes $M_S < M_T$. The similarity of the Gutenberg-Richter statistics for the local region and the surrounding ensemble regions is used to construct a time-scaling for the times in the ensemble cycles.

After time-scaling the calendar forecast times for the ensemble members, we then consider each cycle of activity in a given ensemble. Collections of time-scaled cycles of activity that have fewer small events than the number in the circle are considered to represent a set of might-have-been past histories for activity in the circle. Collections of time-scaled cycles of activity that have more small events than the number in the circle are considered to represent a set of possible future activity for the circle. Examples of the data for a circular region of radius 125 KM around Los Angeles, CA, and a selection of the ensembles is shown in Table 1.

The next step involves computation of the ROC curve for each member of the ensemble, a plot of the True Positive Rate (TPR) vs. the False Positive Rate (FPR) [40,41]. Once we are in possession of these curves, we compute the ensemble mean curve and its standard deviation. The area under the ROC curve (AUROC, or "skill") can be interpreted as the ability of a model to discriminate between, or to correctly classify, different categories of events. In our case, an event is the occurrence of a major earthquake vs. non-occurrence [40,41]. The method then uses a simple data mining method on the longer earthquake cycles to compute the probability of future events.

This current paper is the third of a three-part series introducing methods for computing local earthquake forecasts and associated probabilities of future large earthquakes from local earthquake nowcasts. The first paper [15] detailed a method to compute local forecasts for a fixed future natural time interval, where natural time is the count of small earthquakes since the last large earthquake. The second paper [16] extended these methods to calendar time forecasts using an ensemble-of-regions approach. Here the local region of interest, a circle of radius $r_C$ surrounding a central geographic point of interest, is enclosed by an expanding series (ensemble) of regions.

To summarize our results:

- Local ensemble probabilities for large earthquakes can be computed using only counts of small earthquakes
- Ensemble exceedance probabilities for magnitude and earthquake occurrence time can be computed for evaluating risk profiles

- To test the basic assumption of uniformity of statistics, a nowcast transform can be defined to adjust the regional statistics to enforce uniformity within the ensembles of events
- The method is then validated/benchmarked by comparison to the accepted current USGS forecasts.

# 2 Nowcast Function and the Ensemble Method

We begin by considering a circular region of radius 125 km (**Figure 1a**) surrounding a point of interest, in this case, the city of Los Angeles, CA [5,15,16]. Figure 1a also shows an ensemble of expanding rectangular regions. The last major earthquake within this circle was the magnitude M6.7 Northridge, CA earthquake on January 17, 1994. We wish to compute the probability of the next large earthquake of target magnitude $M_T \geq 6$ occurring within the circular region, whose radius was chosen mainly because it is about twice the radial dimension of the aftershocks of the Northridge earthquake.

## 2.1 Scaled Similarity

These ensemble regions are assumed to posses Gutenberg-Richter magnitude-frequency (GR) statistics that are the same or similar to the statistics in the circular region. This assumption can be stated as an assumption of "scaled similarity", that when the temporal statistics of the members of the ensemble are appropriately scaled in time, the members can be regarded as statistical "copies" of the earthquakes in the circular region.

The specific form of the GR law that is used here is:

$$N = e^{b(M_T - M_S)} \tag{1}$$

where $N$ is the number of small earthquakes of magnitude $M_S$ for every large target earthquake $M_T$, and $b$ is a parameter that is usually near $b \sim 1$ but can typically vary by $\pm$ 0.05 or so.

## 2.2 Step 1: The Nowcast Function and Earthquake Cycles

The first step in the analysis is to introduce a nowcast function $\Phi(n)$:

$$\Phi(n) = 1 - e^{-(n/N)} \tag{2}$$

where $n$ is the number of small earthquakes of magnitude $M_S$ since the last large earthquake of target magnitude $M_T$ [16]. Equation (2) was chosen as a consequence of the observation [42], since validated many times by others [6], that aside from aftershock clustering, earthquakes occur randomly in time with Poisson statistics. In addition, [] has shown that equation (2) generally characterizes the interval statistics (**Figure 1b**) between large earthquakes $M_T$. Note that in the figures, we plot the nowcast value as ranging between [0%,100%] rather than [0,1].

## 2.3 Step 2: Construct Ensemble of Cycles

Using equation (2), build a series of "large earthquake cycles" in the ensembles, shown schematically in **Figure 2**, surrounding the circular region and centered on Los Angeles (**Figure 1a**). Each of the cycles begins and ends at the time of a target earthquake. So an ensemble member with $N_{ENS}$ large earthquakes will have $J = N_{ENS}$ -1 cycles by this definition.

These cycles of activity represent the data mining set for the simple machine learning application.

The $i^{th}$ ensemble member will have a set of $J$ cycles $n_{i,j}$ denoted by $C_i$:

$$C_i\{n_i\} = \{n_{i,1}, n_{i,2}, n_{i,3}, \ldots, n_{i,J}\} \tag{3}$$

Here $n_{i,j}$ is the number of small earthquakes in the $j^{th}$ cycle of the $i^{th}$ ensemble.

## 2.4 Step 3: Scale the Forecast Time for Each Ensemble Member

Build conditional ROC curves for each member of the ensemble using the following method, then compute the mean ROC curve from the ensemble ROC curves. The curves are conditioned on the number of small events that have occurred prior to calculation of the ROC curve.

We choose a forecast time of interest $T_{F,C}$ for the next target earthquake in the circular region. The number of small events in the circle will be designated as $n_C$, and the corresponding rate of small earthquake occurrence within the circle is designated as $R_C$.

Clearly each member of the expanding ensemble will have a larger number of small events than the number of small events in the circle. The $i^{th}$ ensemble member will have a total number $n_i$ of small earthquakes, and a corresponding rate of occurrence $R_i$.

In Step 2, we scale the forecast time interval for each member of the ensemble, $T_{F,i}$ , according to [15,16]:

$$T_{F,i} = \left(\frac{R_C}{R_i}\right) T_{F,C} \tag{4}$$

## 2.5 Step 4: Build Conditional Receiver Operating Characteristic (ROC) Curves

To build the conditional ROC curves for each of the ensembles, we adopt a set of threshold values for the time series amplitudes [10-14]. For each ensemble member, we classify all points on the nowcast timeseries by sweeping the threshold values over all amplitudes of the time series on the interval [0,1].

We start by selecting an arbitrary threshold value $\tau$. Given a small event occurring at time $t$, we ask if nowcast value of that event is above or below the threshold. We also ask if the next target earthquake $M_T$ occurs after time $t$ but within the time interval $t + T_{F,i}$ , where again, $T_{F,i}$ is the scaled forecast time for that ensemble member. Classification is then given by:

- TP, if the nowcast value is above the threshold $\tau$, and the next target earthquake occurs within $t + T_{F,i}$
- FP, if the nowcast value is above the threshold $\tau$, and no earthquake occurs within interval $t + T_{F,i}$
- FN, if the nowcast value is below the threshold $\tau$, and the next target earthquake occurs within interval $t + T_{F,i}$
- TN, if the nowcast value is below the threshold $\tau$, and no target earthquake occurs within interval $t + T_{F,i}$

It should be emphasized that the conditional ROC diagram is conditioned on the current number of events in the circle. For example, if the current number of small earthquakes in the circle is, for example 100, then no cycles of length less than 100 are used in the classification. Examples of conditional ROC diagrams are shown in **Figure 3**. Cyan curves represent the ROC curves for the ensemble members, the red curve is the ensemble mean, and the dashed curves are the standard deviations.

Classifying all points on an ensemble timeseries will then produce a confusion matrix, or contingency table, composed of the quantities TP, TN, FP, FN . We should also explicitly note that these quantities are functions only of the threshold values, $\tau$, so that we have TP($\tau$), TN($\tau$), FP($\tau$), FN($\tau$). Conditional ROC curves for each member of the ensemble are computed, and the mean and standard deviation of the curves are computed. From each confusion matrix, we compute the True Positive Rate (TPR) and the False Positive Rate (FPR):

$$\mathrm{TPR} = \mathrm{TP}/(\mathrm{TP} + \mathrm{FN}) \quad (5)$$

$$\mathrm{FPR} = \mathrm{FP}/(\mathrm{FP} + \mathrm{TN})$$

As discussed, the basic assumption in this method is that the catalog statistics of the surrounding region are the same as the statistics of the circular region. Or more specifically, we make what amounts to an ergodic assumption, that time averages can be replaced by space (or ensemble) averages. In the present paper, we will introduce below a "nowcast transform" that enforces the equality of the GR statistics of the ensemble of regions with the statistics of the circular region.

### 2.6 Step 5: Compute Positive Predictive Value (PPV) from the ROC Curves

Once the conditional ROC curve is computed for a member of the ensemble, the Positive Predictive Value (PPV) is then computed:

$$\mathrm{PPV} = \mathrm{TP}/(\mathrm{TP} + \mathrm{FP}) \quad (6)$$

The PPV value is the probability of a future large earthquake of target magnitude $M_T$ during the forecast interval $T_{F,C}$.

To compute the PPV values (probabilities) at the times of the small events following the last large event in the circle, we use as threshold values $\tau$ the nowcast values $\Phi(n_C)$for the small events in the circle. So if there are 100 small earthquakes within the circle since the last large earthquake, we compute 100 values for $\Phi(n_C)$ and use these as our threshold set $[\tau(n_C)] \equiv [\Phi(n_C)]$. Note that each of these small earthquakes has a calendar time $t$.

Each of these small earthquakes in the circle then has a defined nowcast value, an index, and an occurrence time, all of which are associated with the small event. Again, note that the TP, FP, TN, FN are functions of the threshold only, which in this special case is the event nowcast value. Since the nowcast value is also a threshold, this association allows us to identify specific event sequence numbers and event times with a PPV value. **Figure 4** shows examples of calendar forecasts of this type for both 1 year and 5 year time scales.

## 3 Conditional Exceedance Curves

Here we compute conditional exceedance curves (survivor distributions), conditioned on the number of small earthquakes that have occurred prior to computation of the probabilities.

### 3.1 Step 6: Compute Magnitude Exceedance

To build a magnitude exceedance probability, we note that all the cycles in an ensemble are terminated (and begin) with a magnitude $M \geq M_T$. Thus we can build a set consisting of the next, or terminating, magnitudes $M_{i,j}$ for each interval $n_{i,j}$:

$$C_i\{M_i\} = \{M_{i,1}, M_{i,2}, M_{i,3}, \ldots, M_{i,J}\} \quad (7)$$

Considering both $C_i\{M_i\}$ and $C_i\{n_i\}$, we then build a combined set of intervals and magnitudes:

$$C\{M\} = \cup_{i,j}\{M_{i,1}, M_{i,2}, M_{i,3}, \ldots, M_{i,J}\}$$
$$C\{n\} = \cup_{i,j}\{n_{i,1}, n_{i,2}, n_{i,3}, \ldots, n_{i,J}\} \quad (8)$$

where the $\cup_{i,j}\{\}$ indicates the union of all sets of ensemble intervals and magnitudes into a single large set.

Once we have these two large sets, we can construct the conditional survivor distributions (exceedance probabilities) for terminating magnitudes. Survivor distributions for terminating magnitudes are shown in **Figure 5a**.

Conditional exceedance curves are shown for several examples of natural times (small event counts), for 0 small events (immediately after the last large earthquake), for 448 small events (the current count in the circular region), and for the current-plus-mean-projected number of small events in the circular region. The mean projected number is the average number expected until just before the next large earthquake. This mean projected number is computed by finding the mean of all intervals larger than the current number, 448, minus the current number. Note that a light smoothing has been applied to these curves.

### 3.2 Step 7: Compute Magnitude Exceedance for 25%, 50%, 75% Probability

**Figure 6** analyzes this conditional exceedance data further by computing the 25%, 50% (median), and 75% exceedance probabilities as a function of the number of small events that have occurred. Note the blue vertical line, which indicates the number of small earthquakes that have occurred to date, which is 448. As more small earthquakes occur in the future, the magnitude exceedance probabilities, particularly for the median (50%) and 25% level will begin to increase sharply.

### 3.3 Calendar Time Exceedance

To compute the calendar exceedance, or waiting time until the next large earthquake $M_T$, we first create the set $C(\Delta T_i)$ of combined ensemble time intervals $\Delta T_i$ corresponding to the natural time intervals in equation (8):

$$C\{\Delta T\} = \cup_{i,j}\{\Delta T_{i,1}, \Delta T_{i,2}, \Delta T_{i,3}, \ldots, \Delta T_{i,J}\} \quad (9)$$

Next we scale the time intervals $\Delta T_i$ -> $\Delta T'_i$ using the *inverse* of the scaling factor in equation (4):

$$\widetilde{\Delta T}_{i,j} = \left(\frac{R_i}{R_C}\right) \Delta T_{i,j} \tag{10}$$

leading to:

$$C\{\widetilde{\Delta T}\} = \cup_{i,j}\{\widetilde{\Delta T}_{i,1}, \widetilde{\Delta T}_{i,2}, \widetilde{\Delta T}_{i,3}, \ldots \widetilde{\Delta T}_{i,J}\} \tag{11}$$

Using the sets (8) and (11), we can then construct the conditional exceedance curves for the waiting times, both in natural and calendar times. Results will be discussed below.

## 4 Discussion: Testing the Ensemble Method

Now we test the results of the ensemble method, and explore ways to build ensemble forecasts using two additional ideas. These involve the following:

1. We introduce a "Nowcast Transform" to test the assumption of scaled similarity of the GR statistics in the ensemble.

2. We filter cycle intervals at the 95% confidence level to test an assumption that outliers may determine the statistics.

### 4.1 "Nowcast Transform"

We define this simple method to test the assumption that the statistics of the ensemble members produce a stable forecast result, and to examine the assumption that the statistics of the ensembles being similar to the circle are reasonable.

Using equation (2) we compute the nowcast value for a typical interval in ensemble *i*:

$$\Phi(n_{i,j}) = 1 - e^{-(n_{i,j}/N_i)} \tag{12}$$

where the scale $N_i$ for ensemble member *i* is computed from equation (1):

$$N_i = e^{b_i(M_T - M_S)} \tag{13}$$

and where $b_i$ is the *b*-value for ensemble member *i*.

To compute the transformed intervals $n'_{i,j}$ , we invert equation (2), by making the equivalence $\Phi(n_{i,j}) = \Phi(n'_{i,j})$ , and using the statistics of the circular region:

$$n'_{i,j} = -N_C \; log[1 - \Phi(n_{i,j})] \tag{14}$$

and where $b_C$ and $N_C$ represent the statistics of the circle :

$$N_C = 10^{b_C \,(M_T - M_S)} \tag{15}$$

This process ensures that the transformed ensembles have the same *b*-value, $b_C$, as the circle. Thus the GR interval statistics of the ensembles are the same as those of the circular region.

But in altering the values of the $n_{i,j}$ intervals, we also need to consider the times at which these transformed events occur. There are two cases, one in which $n_{i,j} > n'_{i,j}$ and another in which $n'_{i,j} > n_{i,j}$. We have therefore developed a simple algorithm to assign the small event times for the small events in the intervals:

- For the case $n_{i,j} > n'_{i,j}$ we choose event times in $n_{i,j}$ at random, and remove a sufficient number of these. Continue until the necessary number of times have been removed.
- For the case $n'_{i,j} > n_{i,j}$ we pick two neighboring times randomly, find the mean of these two times, and insert that mean time between the two chosen times. Continue until the necessary number of times have been assigned.

A major feature of this simple algorithm is that the basic structure of the event times is preserved. Where times are densely clustered, as in aftershock intervals, proportionately more times are added or removed. Where times are sparse, as when quiescence prevails, fewer times are added or removed. Thus changes to the temporal density of small event times is proportional to the original density of times. Other algorithms are clearly also possible.

As a final point, since the nowcast transform alters the number of small events in an ensemble, the scaled forecast time must also be further scaled by modifying equation (4) as:

$$T_{F,i} = \left(\frac{R_C}{R'_i}\right) T_{F,C} \tag{16}$$

### 4.2 Filtering the Earthquake Cycles

In the filter method, we consider the set of intervals $C\{n\}$ from equation (8). We compute the mean $\mu$ and standard deviation $\sigma$ of the intervals, and reject any intervals that are larger than a value $\mu + 2\sigma$ corresponding to the 95% confidence limit. From that point, we repeat the analysis from the foregoing.

## 5 Discussion and Examples

### 5.1 Choice of Circular Region

As before, we applied our methods to a circular region of radius 125 km surrounding Los Angeles, CA, using a target magnitude $M_T = 6$ and small earthquake $M_S = 3.49$. The most recent large earthquake in the circle was the Northridge earthquake, having magnitude M=6.7, occurring on January 17, 1994.

The choice of circle radius is of course arbitrary, but we use a rule of thumb that the radius should be at least twice the linear dimension of the aftershock zone of the previous large earthquake in the circle (see Figure 1). In addition, [38] has conducted a more quantitative analysis of appropriate circle radius for similar earthquake magnitudes in Greece, and also arrived at a radius of 125 km.

An additional criterion for the choice of radius is the "radius of significant ground shaking" for a given magnitude, for example the distance at which one might experience Mercalli Intensity VI (PGA $\sim 0.1g$) shaking. This distance is about 100-150 km for an $M_T = 6$ earthquake, which is our minimum size for the forecast [48,49].

### 5.2 Examples of Nowcast and Filtered Calculations

With these considerations, we can now calculate the ROC curves, the PPV calendar time forecasts, and exceedance curves for the original ensembles, the transformed ensembles and for the filtered ensembles.

**Figure 7** shows plots of the PPV curves for the original intervals, the filtered intervals, and the nowcast transformed intervals. The plot for the original intervals (**Figure 7 left)** repeats a plot from **Figure 4** for convenience of comparison. The results show that the final value of PPV for the original intervals are the smallest of the three, which are in order from left to right, 29%, 35%, and 43%. While generally consistent, the three plots should be viewed as alternative views of the forecast probabilities.

**Figure 8** are the corresponding magnitude exceedance curves for the three types of intervals, and **Figure 9** shows the same magnitude exceedances for the 25%, 50% (median), and 75% exceedance probabilities as a function of the number of small events that have occurred. Again, the three sets of plots are generally consistent.

We now compute calendar time exceedance probabilities, the waiting times or lifetimes $\Delta L(t)$, for large earthquake occurrence. To do this, we need to scale the conditional lifetimes since the last large earthquake $M_T$ within each interval $n_{i,j}$. To that end, let the lifetime within an interval $n_{i,j}$ in ensemble $i$ be denoted by $\Delta L(t)_{i,j}$ . Then we scale the lifetimes by the inverse of the factor in equation (4):

$$\Delta L'(t)_{i,j} = \left(\frac{R_i}{R_C}\right) \Delta L(t)_{i,j} \qquad (17)$$

where now $\Delta L'_{i,j}$ is the scaled lifetime since the last large earthquake. We then use the scaled lifetimes to compute the exceedance probabilities.

**Figure 10** shows the comparison for the calendar time exceedance probabilities. In all three cases, we show the Poisson probability curve (dashed line) that is based on the average interval between large earthquakes. Similar to **Figure 5**, we show the curves for several calendar times in the earthquake recurrence cycles. These exceedance curves are for immediately after the last large earthquake, at the current time (Today); at the current time plus 15 years; and at the current time plus 30 years. Note that a light smoothing has been applied to these curves.

A feature of these calendar time curves is that once the initial large earthquake clustering period has passed, the lifetime until the next large earthquake becomes longer and longer as time passes, until the next large earthquake does eventually occur, as noted in [43-47]. This feature, can be inferred by comparing the zero time curve with the Poisson curve. Initially the zero time curve lies below the Poisson curve until a cutoff is reached (shorter waiting times more probable). Subsequently, the zero time curve lies above the Poisson curve (longer waiting times more probable).

We can then repeat the exceedance curve calculation for natural time (small event counts) using equation (8) directly (**Figure 11**). Conditional exceedance curves are shown for several examples of natural times (small event counts): for 0 small events (immediately after the last large earthquake); for 448 small events (the current count in the circular region); and for the current-plus-mean-projected number of small events in the circular region. The mean projected number is the average number of small events expected until just before the next large earthquake. This mean projected number is computed by finding the mean of all intervals larger than the current number, 448, minus the current number. Note that a light smoothing has been applied to these curves.

In this case however, one finds the more intuitively expected result that the longer it has been in natural time since the last large earthquake $M_T$, the shorter the expected natural time until the next large earthquake. Or stated another way, as time since the last large earthquake increases, the probability increases for a shorter lifetime, or time to the next large earthquake.

The reason for the different implications of the calendar and natural time curves is the increasing, nonlinear small earthquake quiescence as time passes. This observation has been discussed in [12]. There it was concluded that the reason for this anomalous slowing down in calendar time (relative to the Poisson curve) is due to an anomalous stiffening/strengthening of the crustal rigidity as the next earthquake approaches, due to the closure of small cracks that, when open, weaken the crust. And as the rigidity of the crust increase, a transition occurs from sliding via unstable stick slip to sliding via stable slip [6,50], as is well known from many laboratory experiments.

# 6 Validating/Benchmarking the Forecasts

An important consideration is to validate the forecasts by comparison to a recognized, accepted forecast or benchmark. The obvious choices are the UCERF2 time-independent forecast, and the UCERF3 time-dependent forecast [51,52]. **Figure 12** shows the comparison. The UCERF2 forecast for magnitude 6.7, produced in 2007, computed a 30 year forecast of 67% for the Los Angeles box, a rectangular region that we show in the new Figure 5, and a 63% for the San Francisco Bay area for that rectangular box. In 2016, the UCERF3 forecast published the 30 year forecast of 60% for the same Los Angeles box. The UCERF2 30-yr forecast is especially relevant because we are past the midpoint of the 30-year time window. The UCERF3 time-dependent forecast is a valid comparison because it is the official USGS forecast.

The UCERF3 forecast was produced in 2014, whereas the ensemble forecast was produced now. But the numbers are similar, and within the quoted error bounds for the ensemble method. This similarity provides a useful benchmark for evaluating the quality of the ensemble forecast.

# 7 Concluding Remarks

We have developed a new method for earthquake forecasting for arbitrary forecast times and for small geographic regions encompassing multiple earthquake faults. It is generally not as appropriate for individual faults, where quiescence may be the normal mode of behavior. Since the forecasts are conducted in regions, rather than on individual faults, stability of the Gutenberg-Richter statistics and the associated $b$-value are observed using least squares fits to the magnitude-frequency curves [53]. The assumption of similar Gutenberg-Richter statistics also amounts to the assumption that we confine analysis to a single tectonic province [53].

The method is appropriate for use when a forecast for such small regions is desired on a frequent basis, and can be carried out rapidly and easily on a laptop computer in a few minutes of computation. For that reason, the ensemble method constitutes a convenient addition to present forecast methods.

**Author Contributions.** Conceptualization, JBR, IB, AD, LGL, GF and KN; methodology, JBR and KN; software, JBR and KN; validation, JBR, IB and KN; formal analysis, JBR; investigation, JBR; resources, JBR and IB; data curation, JBR; writing—original draft preparation, JBR; writing—review and editing, AD, LGL, GF and KN; visualization, JBR; supervision, JBR; project administration, JBR.; funding acquisition, JBR. All authors have read and agreed to the published version of the manuscript."

**Funding.** Funding for this project has been provided by a generous gift from Dr. John LaBrecque to the University of California, Davis.

**Data Availability Statement.** Data for this paper was downloaded from the USGS earthquake catalog for California, and are freely available there. An included method in the Python code mentioned above can be used to download these data for analysis. Python code that can be used to reproduce the results of this paper can be found at the Zenodo site: https://doi.org/10.5281/zenodo.19390594

**Acknowledgements**. The authors would also like to acknowledge an informative conversation with Jeanne Hardebeck of the USGS.

# 8 Tables

**Table 1.** Statistical data for selected ensemble members for cycles between successive magnitude $M \geq 6$ earthquakes from 1980.0-present. Min/Max Cycle Length is the minimum/maximum number of small earthquakes among the cycles for that region size. Total Small EQ entry is the total number of small earthquakes in the catalog for the circle or ensemble member since 1980 for the range $M \geq 3.5$. Least squares fits in each ensemble member are used to compute the Gutenberg-Richter $b$-values. The range of magnitude values used for computing the $b$-values are between $M = 3.5$ and $M = 6.5$.

| | | Ensemble Number | | | | | | |
|---|---|---|---|---|---|---|---|---|
| | | 1 | 5 | 10 | 15 | 20 | 25 | 30 |
| Region Size | 125 Km LA Circle | 3.6° x 3.6° Rectangle | 4.0° x 4.0° Rectangle | 4.5° x 4.5° Rectangle | 5.0° x 5.0° Rectangle | 5.5° x 5.5° Rectangle | 6.0° x 6.0° Rectangle | 6.5° x 6.5° Rectangle |
| Total Small EQ | 600 | 6286 | 6657 | 7345 | 7803 | 8026 | 8192 | 8777 |
| Number Cycles | - | 22 | 24 | 27 | 29 | 32 | 33 | 41 |
| Min Cycle Length | - | 15 | 10 | 10 | 10 | 2 | 6 | 7 |
| Max Cycle Length | - | 657 | 797 | 800 | 825 | 730 | 755 | 798 |
| b-value | $0.93 \pm .02$ | $0.96 \pm .01$ | $0.96 \pm .01$ | $0.97 \pm .01$ | $0.98 \pm .01$ | $0.96 \pm .01$ | $0.97 \pm .01$ | $0.94 \pm .01$ |

# 9 Figure Captions

**Figure 1. Left**: Regional seismicity (small dots) used in this paper. Large earthquakes are shown as red circles. A region of radius 125 KM around the city of Los Angeles is shown as a blue circle. **Right:** Green bars show the histogram of the number of small earthquakes between large earthquakes. Solid red stair-step line is the Cumulative Distribution Function (CDF) corresponding to the histogram. Magenta dashed lines are the standard deviations from the CDF, computed using a bootstrap method. Red dot is the current count (= 448) of small earthquakes in the circular region since the January 17, 1994 Magnitude 6.7 Northridge, CA earthquake. Current count corresponds to a CDF/Nowcast Poisson value of 94.1%. Blue dashed line is the Poisson CDF, which we use as the "Nowcast Function", and is based on the average (Gutenberg Richter) number of small earthquakes in the region (see equation (1) in the text).

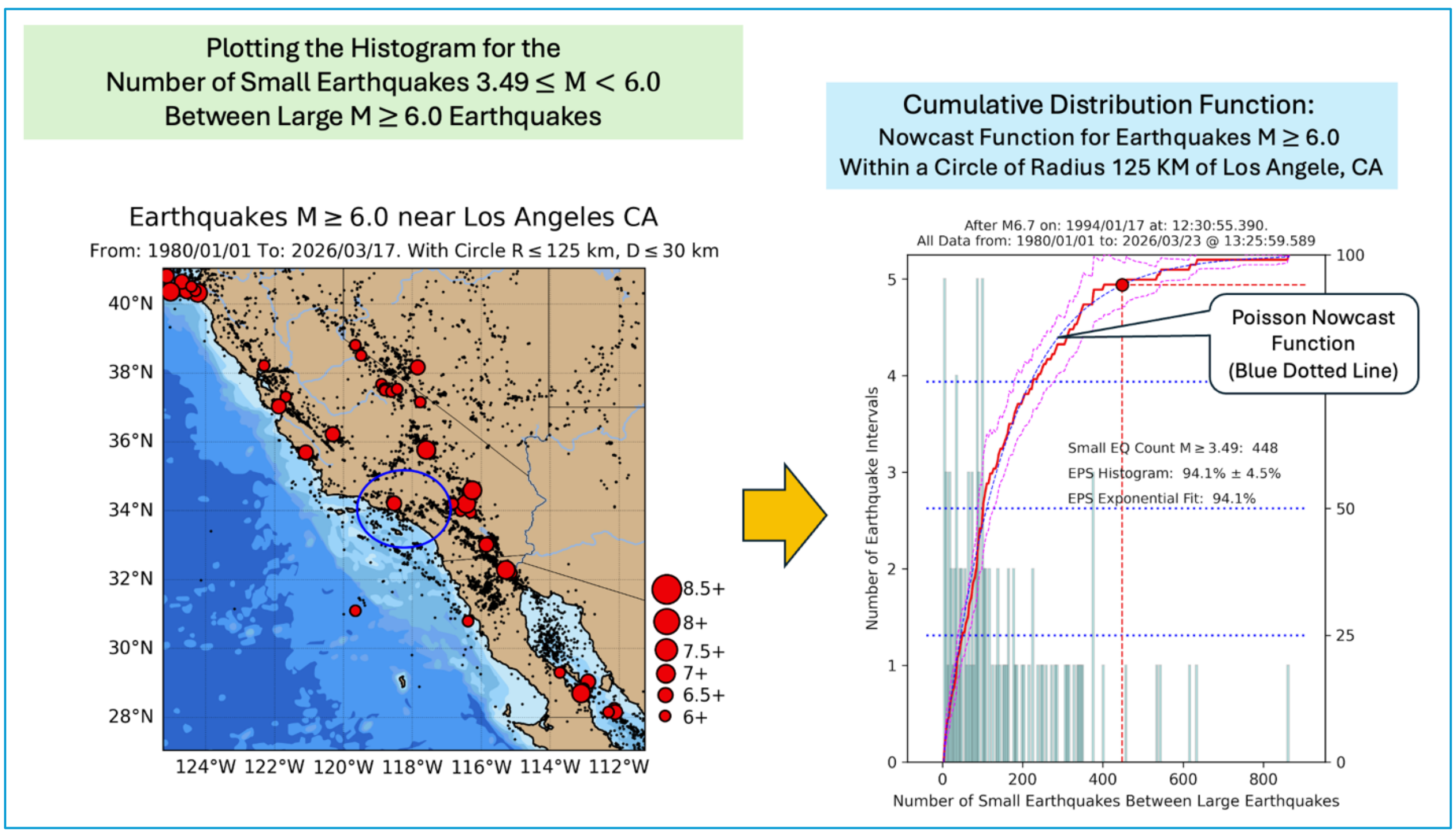

**Figure 2.** Schematic illustration of the ensemble method. **Left:** A series of expanding regions (white rectangles) are constructed surrounding the circular region. **Right:** Using the nowcast function (equation (2)) a time series of nowcast earthquake cycles for each region in the ensemble is constructed using the small earthquakes $3.49 < M_S < M_T$ between the large target earthquakes $M_T \geq 6.0$.

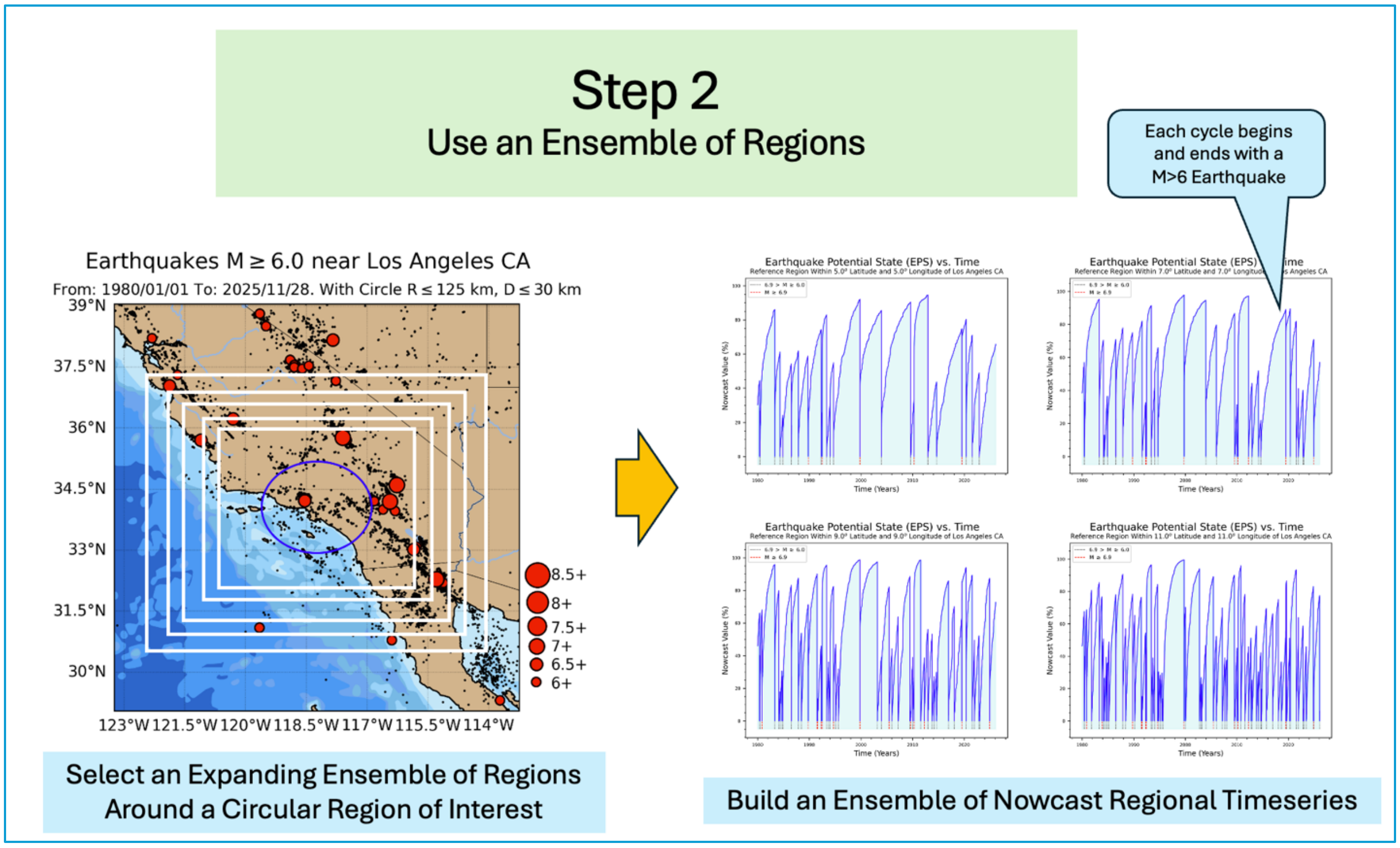

**Figure 3.** Conditional Receiver Operating Characteristic (ROC) diagrams for an ensemble of 30 regions from 3.6º to 6.5º surrounding Los Angeles, CA, at 0.1º interval half-widths, with a forecast time of $T_F$ = 5 years. **a)** ROC diagram computed after no small earthquakes have occurred. **b)** ROC diagram computed after 150 small earthquakes have occurred. **c)** ROC diagram computed after 300 small earthquakes have occurred. **d)** ROC diagram computed after 448 small earthquakes have occurred, which is the number in the circle to date. The diagrams are conditional because only cycles with more events than the designated number (0, 150, 300, 448) of events are used to compute the ROC curves. Cyan curves are for the various ensemble members. Red curve is the mean value, dashed curves are the 1 standard deviation curves. Diagonal line is the no skill line. Skill for the curves are, respectively, 0.47, 0.78, 0.86, 0.90, showing that skill improves progressively as the earthquake cycle proceeds.

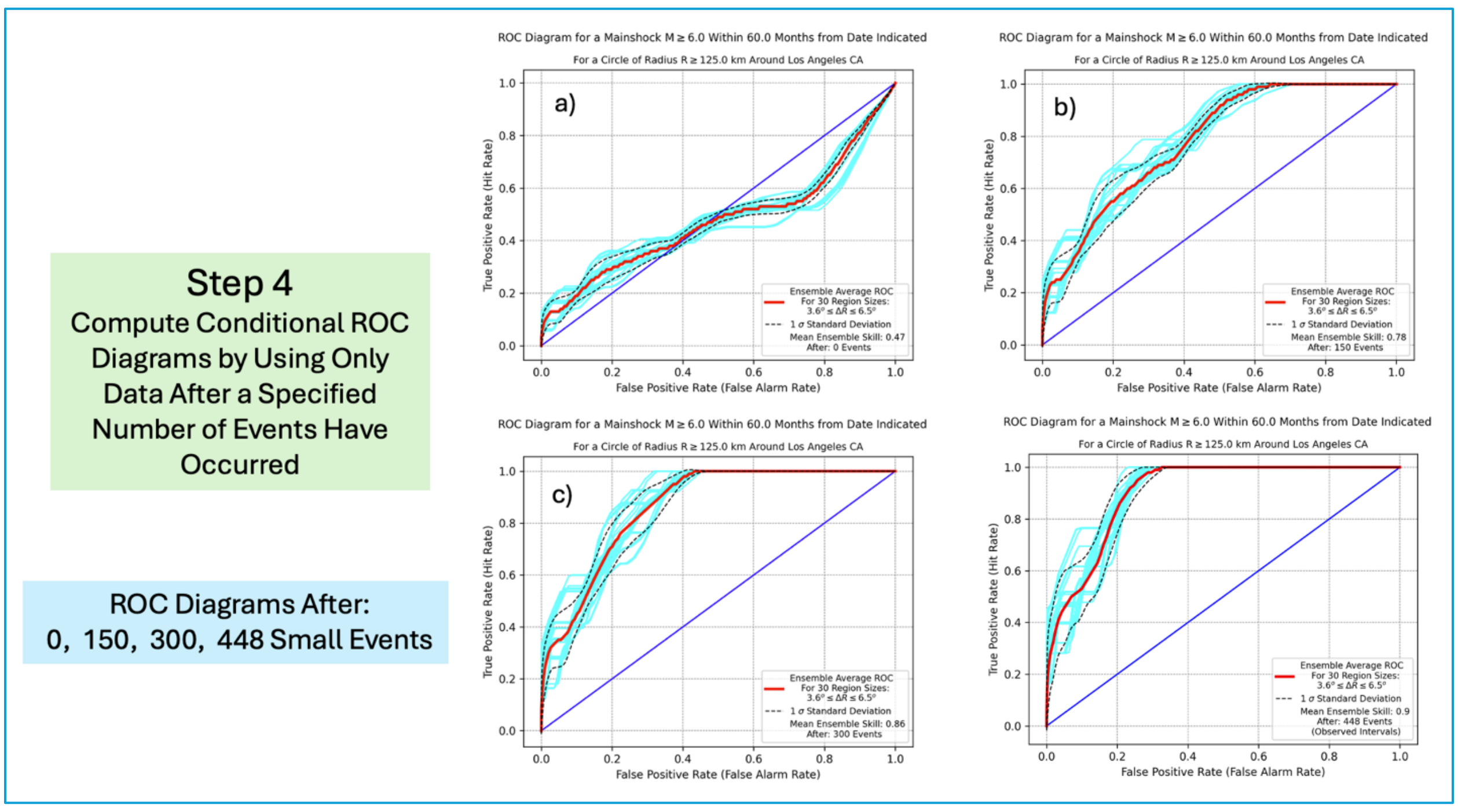

**Figure 4.** Plots of PPV, the probability of a future $M \geq 6$ earthquake, as function of time since the M6.7 Northridge, CA earthquake on 1/17/1994. **a)** Ensemble size = 30, Forecast time interval $T_F$ = 1 year. **b)** Ensemble size = 60, Forecast time interval $T_F$ = 1 year. **c)** Ensemble size = 30, Forecast time interval $T_F$ = 5 years. **d)** Ensemble size = 60, Forecast time interval $T_F$ = 5 years. Cyan curves are for the various ensemble members. Red curve is the mean of the cyan curves, the ensemble probability, and dashed curves are the 1-standard deviation curves. In figures **a)** and **b)**, probability immediately after the mainshock is high, indicating a tendency for mainshock clustering. In figures **c)** and **d)**, tectonic reloading over time following the Northridge earthquake is the primary process that can be seen.

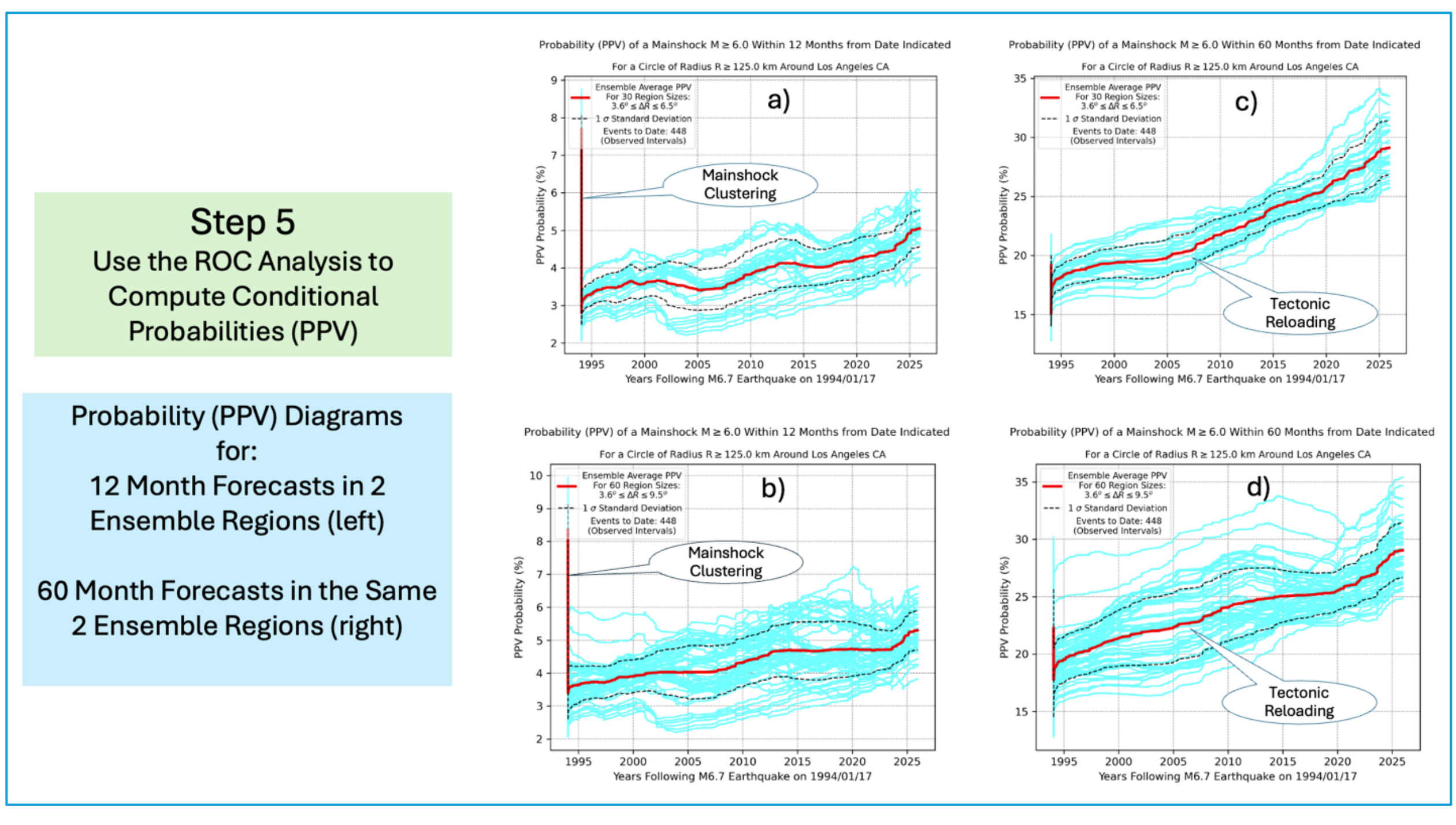

**Figure 5.** Conditional Exceedance Probabilities for magnitude of the terminating target earthquake $M \geq M_T$, after a number of small earthquakes have occurred. **a)** Base case, just after the last large earthquake, in which no small earthquakes have yet occurred. **b)** Current case for number of small earthquakes (448) that have occurred. **c)** A future time at which 615 small earthquakes have occurred.

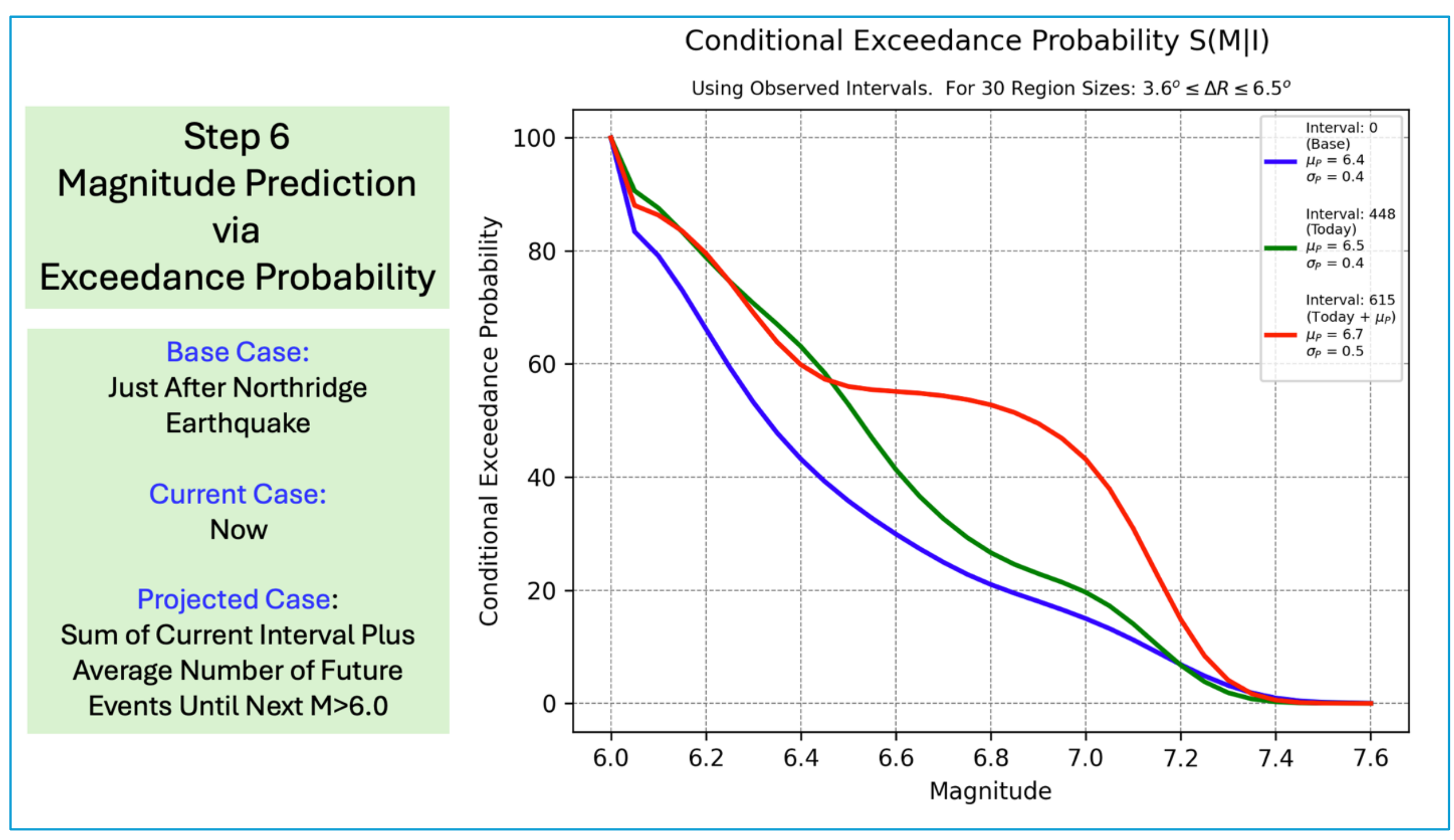

**Figure 6.** Plot of expected terminating magnitude vs. natural time (small event counts) for 25%, 50%, and 75% values of exceedance probability.

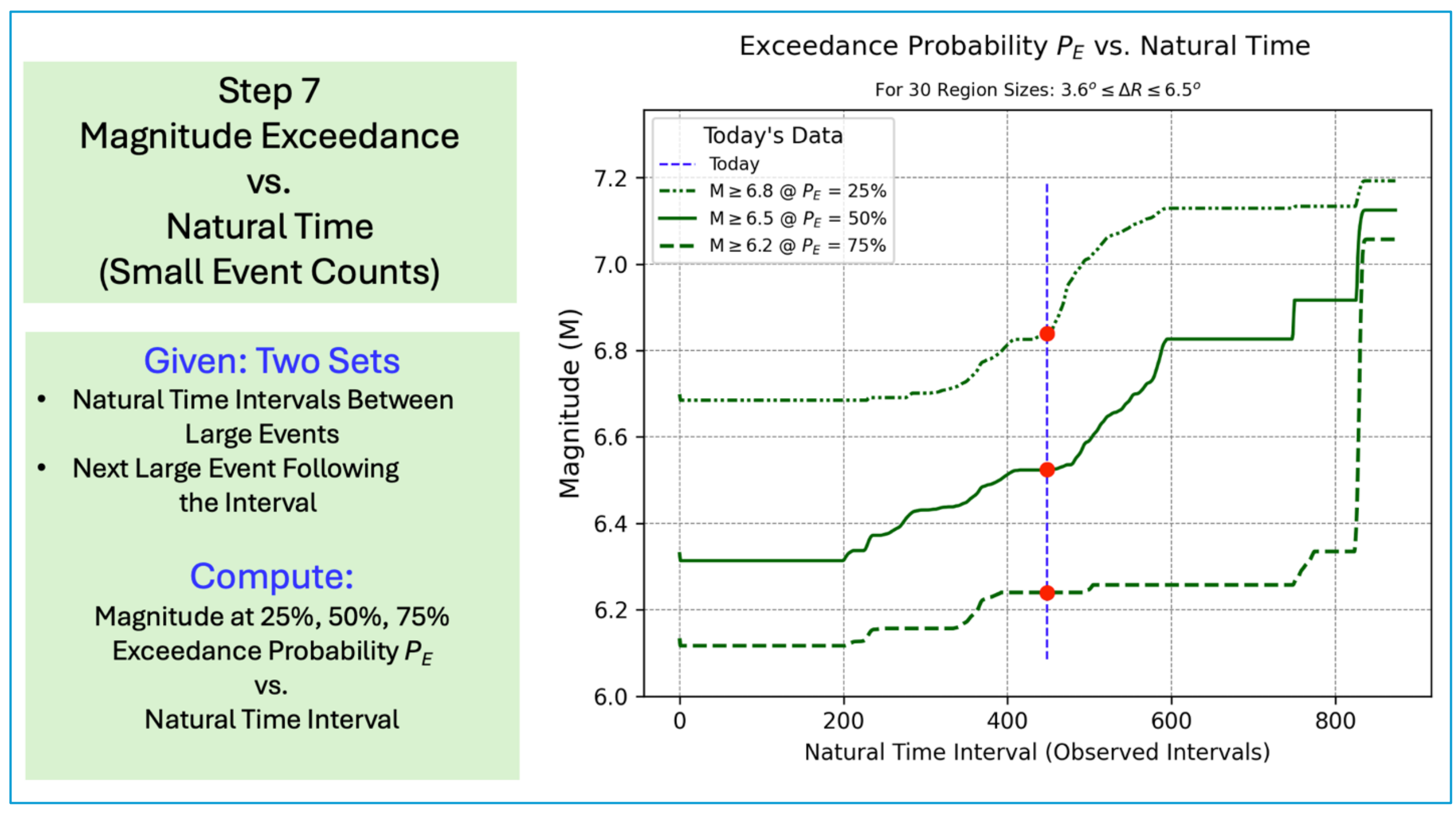

**Figure 7.** Comparison of PPV vs. calendar time forecasts for original observed intervals, filtered intervals, and transformed intervals, for a 60 month (5 year) period using ensembles of size 30 members.

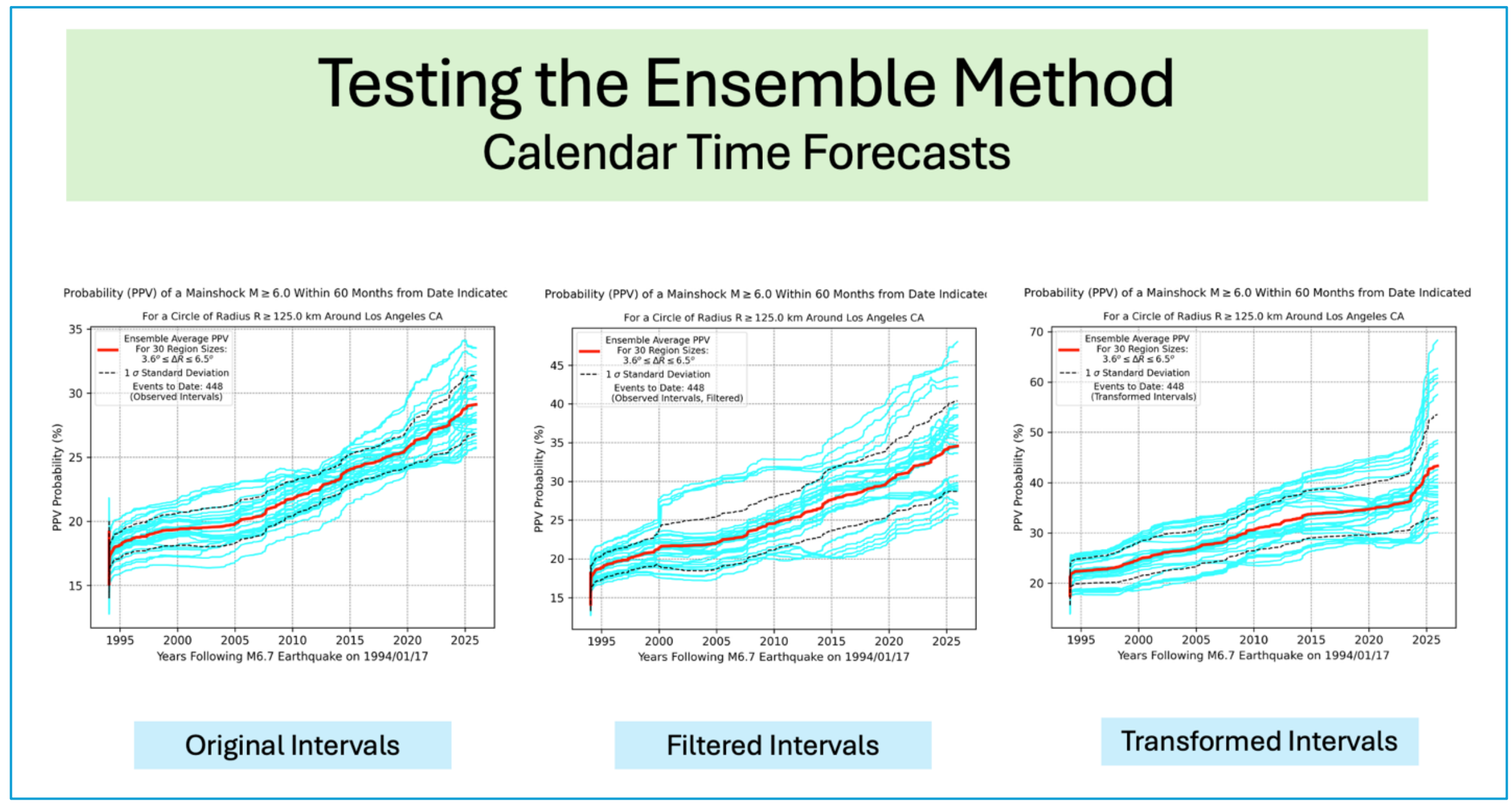

**Figure 8.** Terminating magnitude exceedance probabilities for original observed intervals, filtered intervals, and transformed intervals, using 30 ensemble members.

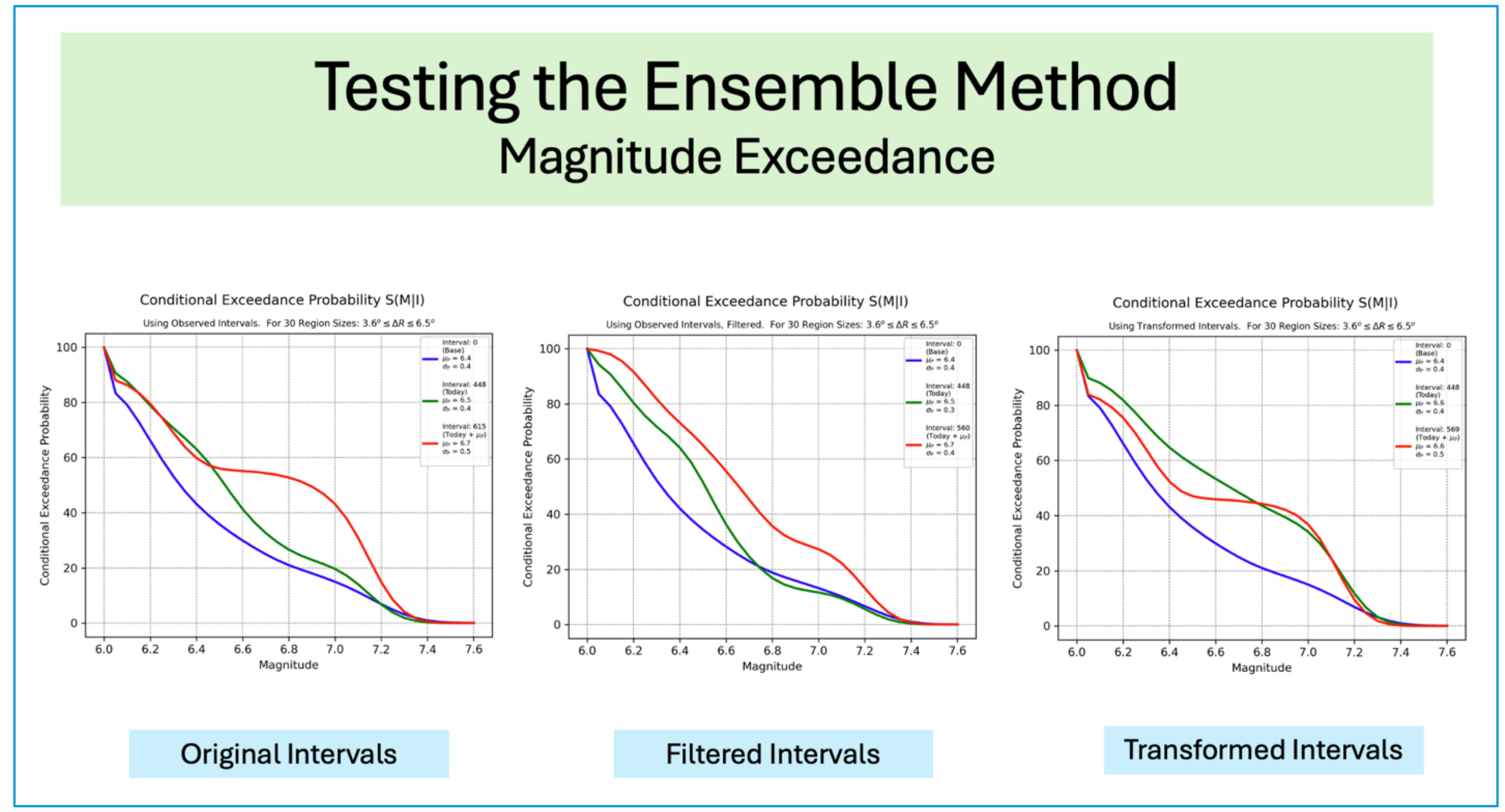

**Figure 9.** Magnitude exceedance probabilities vs. natural time for original observed intervals, filtered intervals, and transformed intervals for 25%, 50%, and 75% values of exceedance probability.

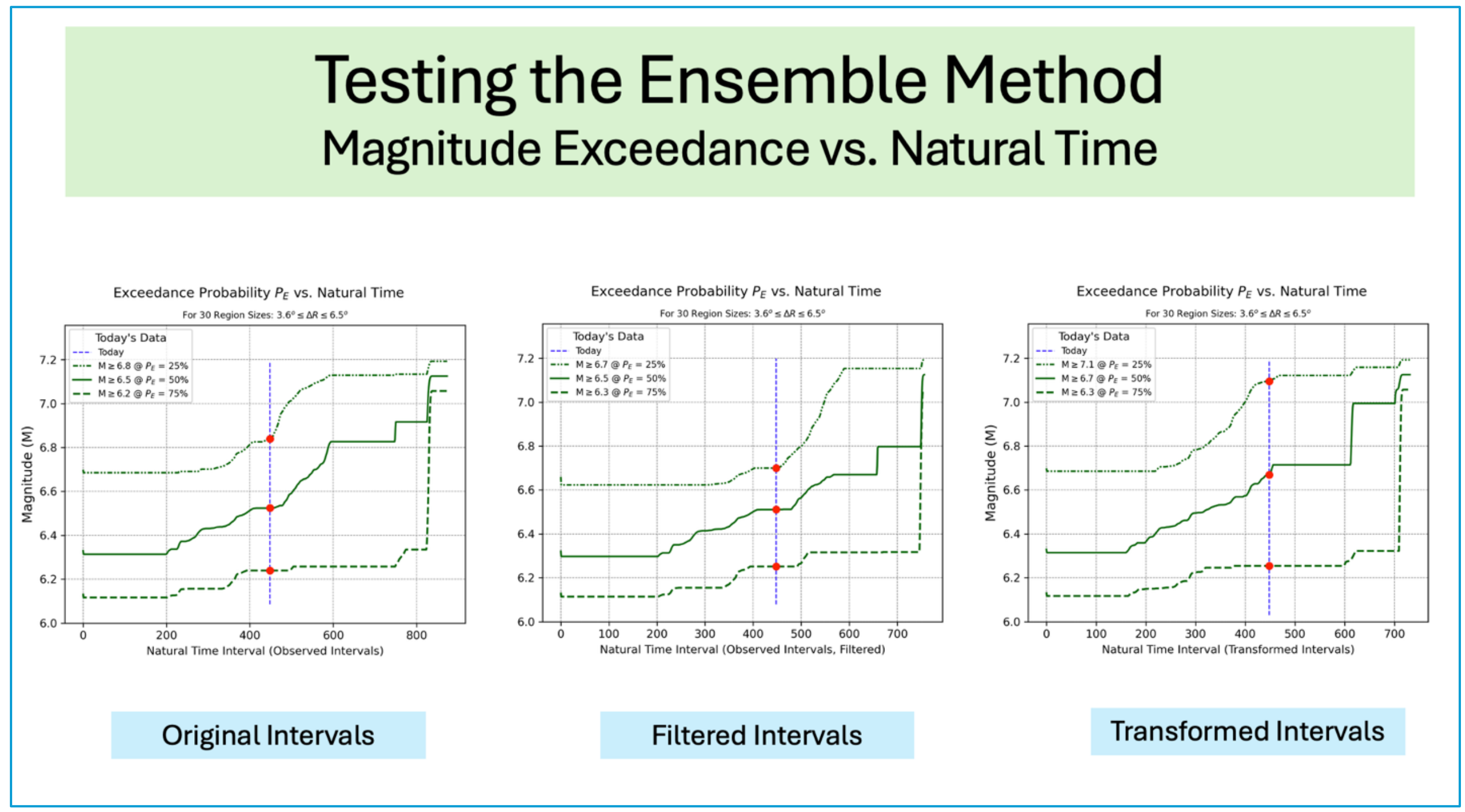

**Figure 10.** Calendar time exceedance probabilities for original observed intervals, filtered intervals, and transformed intervals. Conditional curves are shown for elapsed times (following the last large earthquake) of 0 years; Today, 31.91 years, the current elapsed time following the 1994 Northridge earthquake; Today + 15 years from now; and Today + 30 years from now.

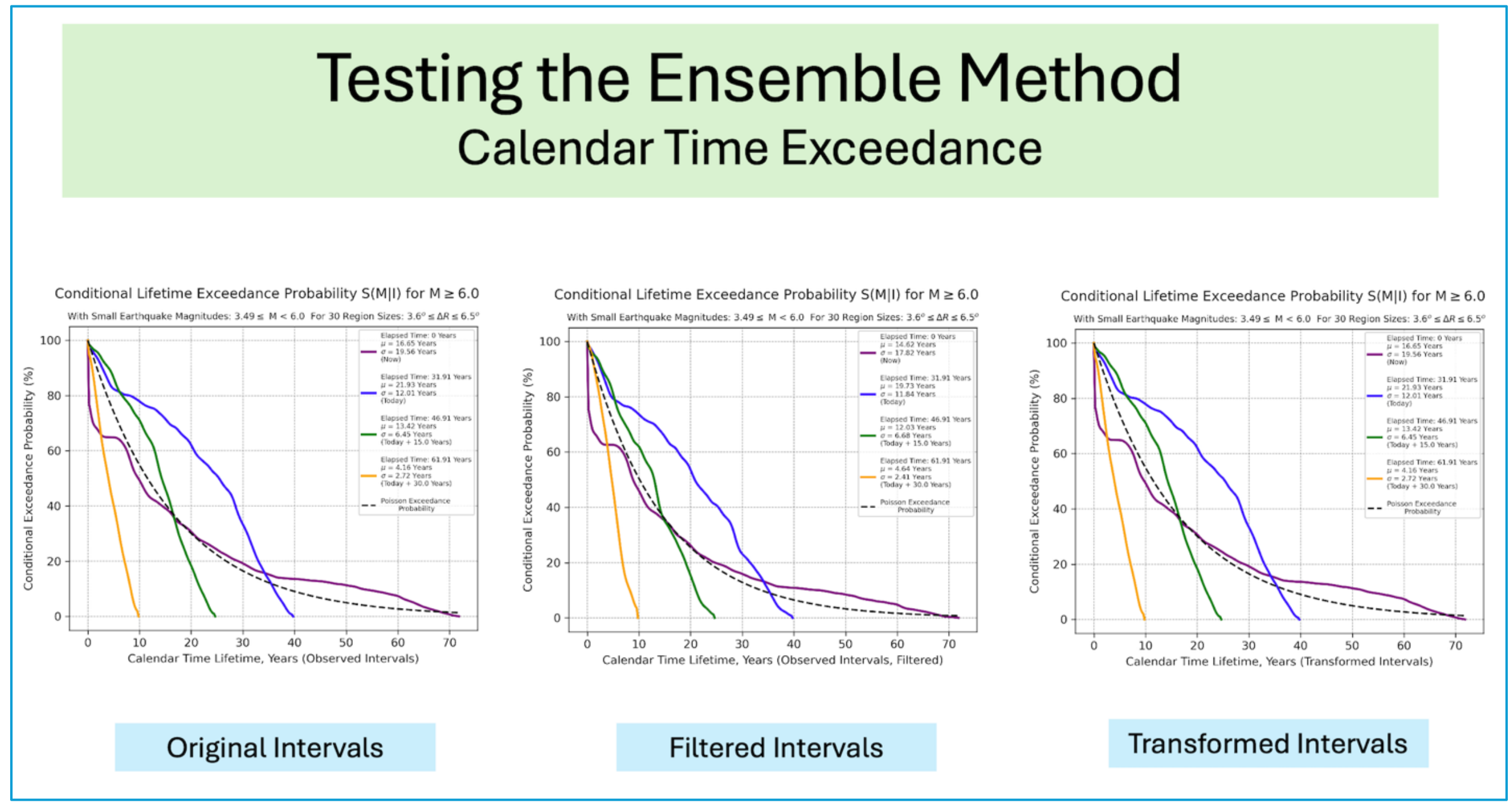

**Figure 11.** Natural time exceedance probabilities for original observed intervals, filtered intervals, and transformed intervals. Conditional curves are shown for elapsed natural times (following the last large earthquake) of 0 small events; Today, 448 small events, the current number following the 1994 Northridge earthquake; Today + 150 additional small earthquakes; and Today + 300 additional small events.

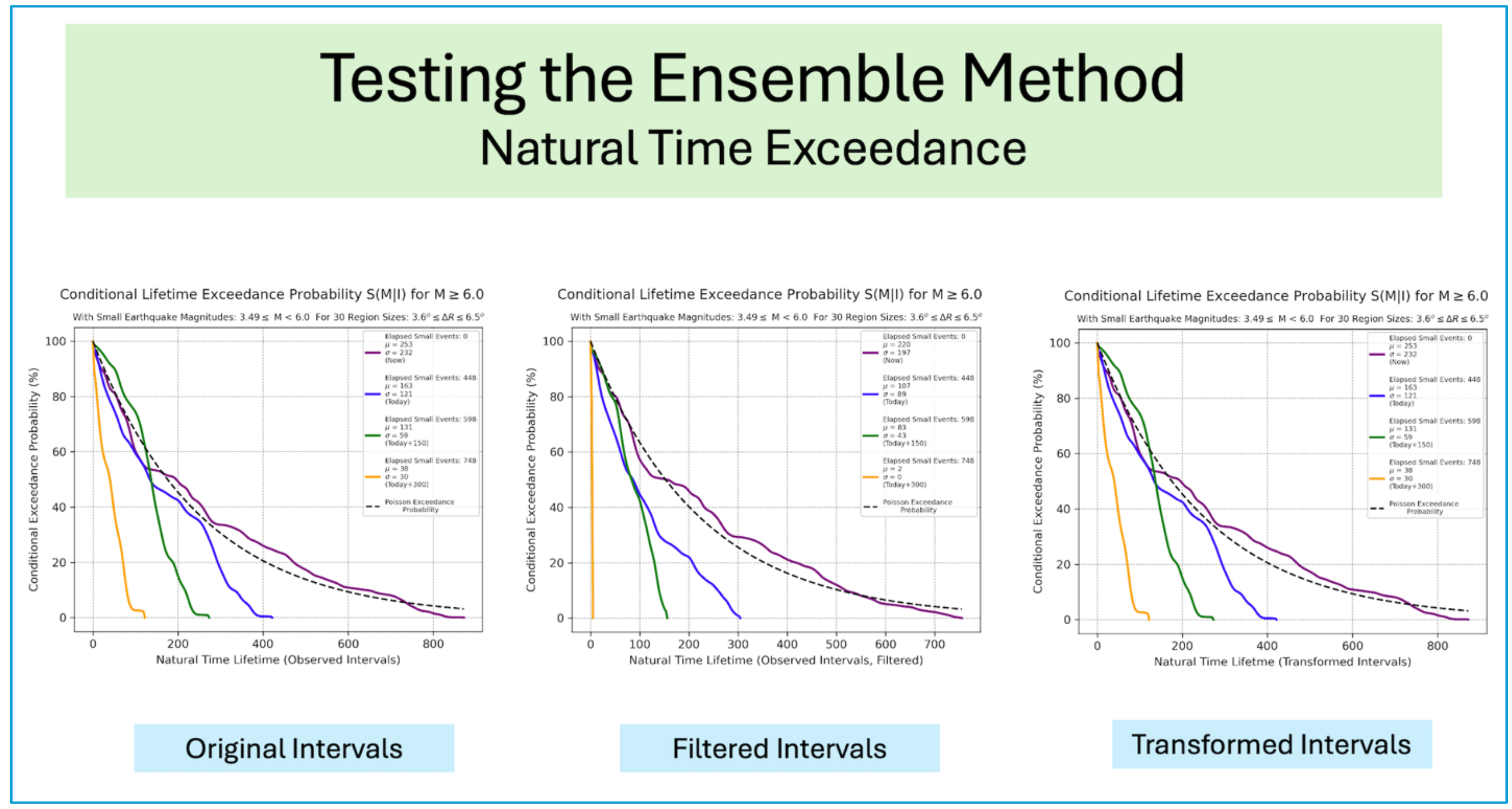

**Figure 12.** Validation of the ensemble method by comparing to the UCERF3 time-dependent forecast (Field et al., 2014) for the Los Angeles and San Francisco boxes defined in the UCERF3 forecast. The forecast curves at right begin at the time of the 1994 Northridge earthquake for the Los Angeles box, and the 1989 Loma Prieta earthquake for the San Francisco box, respectively. The current values of the Los Angeles and San Francisco ensemble 30-year forecasts are listed in the figure, together with the UCERF3 30-year forecasts.

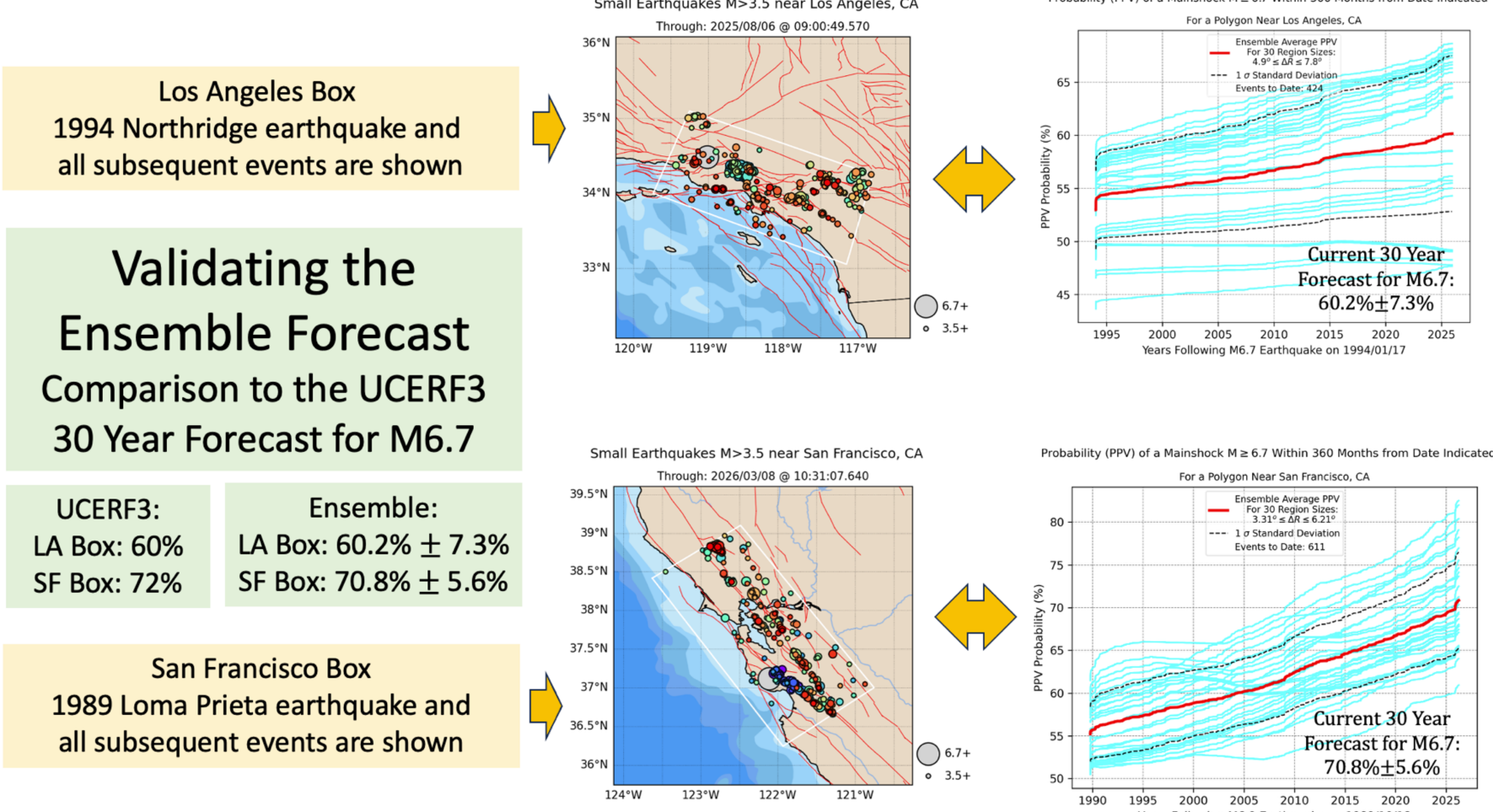